\documentclass[aps,prb,reprint,amsmath,amssymb,showpacs,superscriptaddress,notitlepage,longbibliography]{revtex4-2}
\usepackage[colorlinks=true,linkcolor=blue,anchorcolor=red,citecolor=blue,urlcolor=blue]{hyperref}
\usepackage{graphicx}
\usepackage{amsfonts}
\usepackage{caption}
\usepackage{ragged2e}
\usepackage{microtype}
\usepackage{amsmath}
\usepackage{amssymb}
\usepackage{bm}
\usepackage{color}
\usepackage{float}
\DeclareCaptionJustification{fullyjustified}{\justifying}
\newcommand{\frameworkimage}[5][\linewidth]{%
  \IfFileExists{#2}{\includegraphics[width=#1]{#2}}{%
    \IfFileExists{#3}{\includegraphics[width=#1]{#3}}{%
      \fbox{\parbox[c][#4][c]{0.92\linewidth}{\centering\small #5}}}}}

\begin{document}
\title{Magnetic-Field Response of a Hyperbolic Kitaev Spin Liquid}

\author{Tao Yang}
\affiliation{National Laboratory of Solid State Microstructures and Department of Physics, Nanjing University, Nanjing 210093, China}

\author{Baigeng Wang}
\email{bgwang@nju.edu.cn}
\affiliation{National Laboratory of Solid State Microstructures and Department of Physics, Nanjing University, Nanjing 210093, China}
\affiliation{Collaborative Innovation Center of Advanced Microstructures, Nanjing University, Nanjing 210093, China}
\affiliation{Jiangsu Physical Science Research Center, Nanjing University, Nanjing 210093, China}

\author{Rui Wang}
\email{rwang89@nju.edu.cn}
\affiliation{National Laboratory of Solid State Microstructures and Department of Physics, Nanjing University, Nanjing 210093, China}
\affiliation{Collaborative Innovation Center of Advanced Microstructures, Nanjing University, Nanjing 210093, China}
\affiliation{Jiangsu Physical Science Research Center, Nanjing University, Nanjing 210093, China}
\affiliation{Hefei National Laboratory, Hefei 230088, People's Republic of China }

\author{Z. Y. Xie}
\email{qingtaoxie@ruc.edu.cn}
\affiliation{School of Physics, Renmin University of China, Beijing 100872, China}
\affiliation{Key Laboratory of Quantum State Construction and Manipulation (Ministry of Education), Renmin University of China, Beijing 100872, China}

\def\tred#1{{\textcolor{red}{#1}}}
\captionsetup[figure]{name=Fig.}
\renewcommand{\thefigure}{\arabic{figure}}
\renewcommand{\thetable}{\arabic{table}}
\renewcommand{\theequation}{\arabic{equation}}

\begin{abstract}
Hyperbolic geometry can give rise to Kitaev spin liquids with distinct flux structures and Majorana spectra, raising the question of how these fractionalized states respond beyond exactly solvable limits. We address this question on the Kekul\'e-colored hyperbolic $\{8,3\}$ lattice by developing an infinite projected entangled-pair-state approach with
flux-resolved loop-gas initialization and geometry-adapted tensor-network contraction.
Under uniform and staggered $[111]$ magnetic fields, we find markedly different many-body responses.
The uniform field induces a sharp change between
competing variational branches at weak field, whereas
the polarization and mean plaquette flux evolve over
a much broader field range under the staggered field.
Exchange-energy competition helps explain the different
response scales, while third-order effective Majorana
models in a fixed-flux background exhibit contrasting
low-energy spectra.
Our results connect the geometry-dependent fractionalized
structure of a solvable hyperbolic spin liquid to its
many-body response under physical perturbations that
break flux conservation.
\end{abstract}

\maketitle

\section{Introduction}
\label{sec:introduction}

The Kitaev honeycomb model showed how bond-dependent spin exchange can give
rise to fractionalized excitations and topological order~\cite{Kitaev2006,ZhouKanodaNg2017,MatsudaShibauchiKee2025}. Its exact
solution describes itinerant Majorana fermions coupled to an emergent,
static $\mathbb{Z}_2$ gauge field~\cite{Kitaev2006}. The Majorana
construction extends to other lattices with one $x$, one $y$, and one $z$
bond at each site~\cite{YaoKivelson2007,MandalSurendran2009}. Keeping this local exchange rule fixed lets us ask how
the choice of lattice shapes the flux background and the Majorana
spectrum. Hyperbolic $\{p,3\}$ lattices offer a natural setting for this
question, retaining threefold coordination while introducing negative
curvature and noncommuting translations~\cite{Boettcher2022,MaciejkoRayan2022,HPL2025}.

On the honeycomb lattice, the ground state is flux free and, at equal
exchange couplings, has a Dirac Majorana spectrum with a vanishing
zero-energy density of states~\cite{Kitaev2006}.
The flux background and Majorana spectrum can differ markedly on
hyperbolic $\{p,3\}$ lattices~\cite{MosseriEtAl2025,DuselEtAl2025,VidalMosseri2025}.
In a particular reflection-symmetric realization, the ground state
carries uniform $\pi$ flux, and the isotropic Majorana spectrum
realizes a Majorana metal with finite zero-energy density of
states~\cite{HPL2025}.
The low-energy spectral weight is dominated
by non-Abelian Bloch states associated with higher-dimensional
representations of the hyperbolic translation group~\cite{HPL2025,LenggenhagerSupercells2023}.
These results illustrate how geometry can reshape Kitaev
spin-liquid physics, while leaving its response beyond the
exactly solvable limit open.

\begin{figure}[htbp]
    \centering
    \includegraphics[width=0.8\columnwidth]{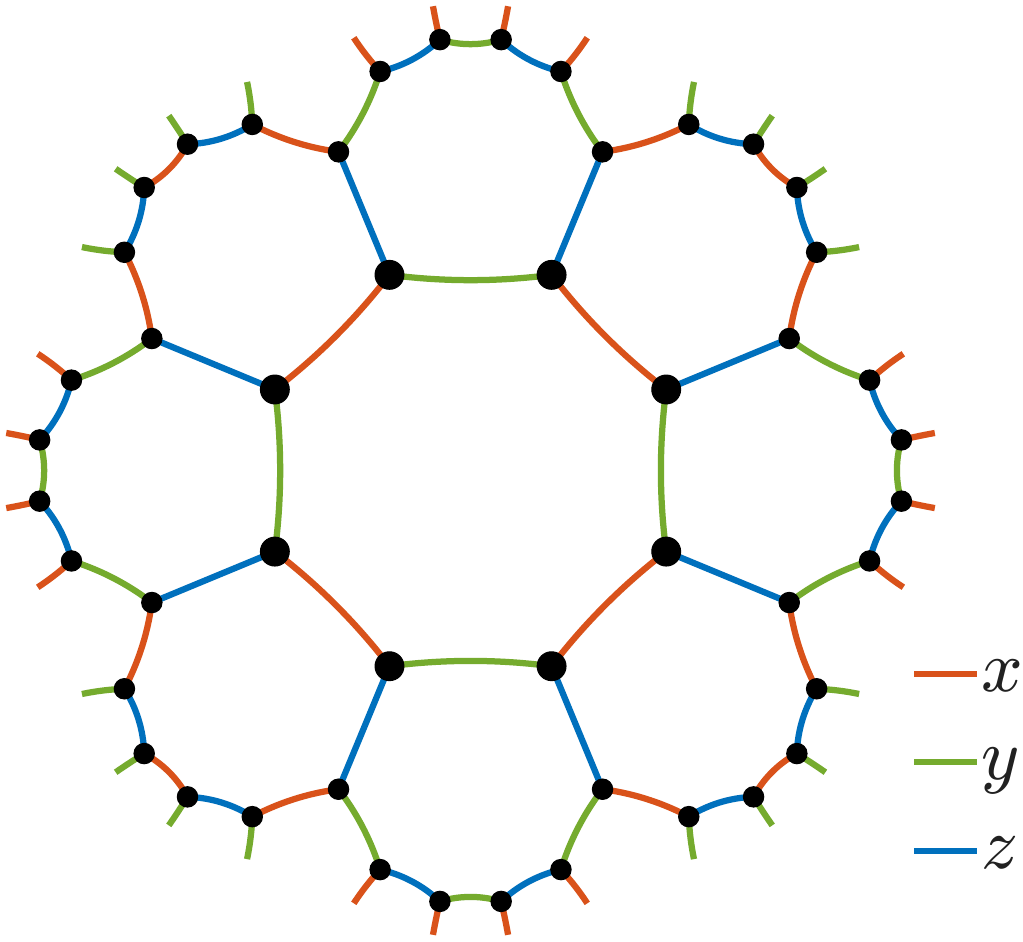}
    \caption{
    Hyperbolic $\{8,3\}$ lattice with Kekul\'e bond coloring.
    Red, green, and blue bonds correspond to the $x$, $y$, and $z$
    Kitaev interactions, respectively.
    }
    \label{fig:hyperbolic_lattice}
\end{figure}

Studying such responses connects the Kitaev problem to the broader study
of correlated quantum matter on hyperbolic lattices. Synthetic circuit
realizations and hyperbolic band theory have opened routes to studying
non-Euclidean quantum physics~\cite{Kollar2019,MaciejkoRayan2021,LenggenhagerCircuit2022}.
Studies of the Hubbard and Heisenberg models have further revealed
geometry-dependent magnetic phenomena and marked differences between
bulk and boundary responses~\cite{Gotz2024,HickeyMaciejko2026}. A central numerical
challenge is that boundary sites remain a finite fraction of a growing
open hyperbolic cluster, so cluster-wide averages can retain substantial
boundary contributions~\cite{LeeBoundary2016,Gotz2024}.

To study the Kitaev spin liquid beyond the solvable limit,
we develop an infinite projected entangled-pair state (iPEPS) approach~\cite{VerstraeteCirac2004,JordanEtAl2008} directly on the
hyperbolic $\{8,3\}$ lattice, whose geometry and
Kekul\'e bond coloring are illustrated in
Fig.~\ref{fig:hyperbolic_lattice}. To facilitate the optimization of strongly
entangled spin-liquid states, we initialize imaginary-time
evolution with flux-resolved loop-gas wave functions~\cite{LG2019}.
For the tensor-network contraction, we develop a
geometry-adapted corner transfer matrix renormalization
group (GA-CTMRG) that respects the native hyperbolic
connectivity and enables the evaluation of local bulk
observables without relying on open-cluster averages.
Together, these ingredients allow us to study competing
zero-field states and their response to magnetic fields.


A physical Zeeman field mixes plaquette-flux sectors,
taking the Kitaev spin liquid beyond its exactly solvable
Majorana description~\cite{Kitaev2006,GohlkeEtAl2018}.
We compare uniform and staggered $[111]$ fields of equal
local magnitude to explore how the spatial structure of
a perturbation affects this fractionalized state.
For ferromagnetic Kitaev exchange, uniform polarization
is compatible with the preferred bond alignment, whereas
staggered polarization competes with it.
The two field patterns also preserve different symmetries,
which constrain their possible low-energy responses.
We examine how these distinctions manifest in polarization
and plaquette flux, as well as in the weak-field effective
Majorana spectra.

At zero field, our optimized iPEPS favor the
$\pi$-flux-initialized branch at larger bond dimensions,
consistent with the established ground-state flux.
Exchange anisotropy redistributes the bond correlations
while preserving the uniform $\pi$-flux background.
Taking the isotropic spin liquid as a common starting
point, we find markedly different responses to uniform
and staggered $[111]$ magnetic fields.
Under the uniform field, the energetically favored
variational solution changes within a narrow weak-field
range, accompanied by a sharp increase in polarization
and a mean plaquette flux approaching zero.
In contrast, both observables evolve over a much broader
field range under the staggered field, without a
comparably sharp change.
The distinct response scales can be understood in part
from the different exchange-energy costs of uniform and
staggered polarization.
A complementary distinction emerges in the fixed-$\pi$-flux
third-order effective Majorana models: the uniform-field
perturbation retains low-energy spectral weight, whereas
the staggered-field perturbation depletes it.
Together, these results connect the field-driven many-body
response to the exchange energetics and fractionalized
low-energy physics of the hyperbolic Kitaev spin liquid.

\section{Hyperbolic Kitaev model and flux-resolved tensor-network states}
\label{sec:model_states}

\subsection{Lattice, Hamiltonian, and plaquette fluxes}
\label{subsec:model_flux}

The Kitaev Hamiltonian on the Kekul\'e-colored $\{8,3\}$ lattice is
\begin{equation}
H_K=-\sum_{\langle rr'\rangle_\alpha}
J_\alpha\sigma_r^\alpha\sigma_{r'}^\alpha,
\label{eq:HK}
\end{equation}
where $\sigma_r^\alpha$ are Pauli operators and $J_\alpha>0$ denotes
ferromagnetic exchange. Each site is connected to one $x$, one $y$,
and one $z$ bond. The octagonal plaquettes fall into three classes,
whose boundaries alternate between $(y,z)$, $(z,x)$, and $(x,y)$ bonds,
respectively [Fig.~\ref{fig:loop_gas}(b)]. This coloring is the
hyperbolic analogue of the honeycomb Kekul\'e bond pattern and
preserves reflections through bonds~\cite{HPL2025,VidalMosseri2025}.

For a plaquette of type $\gamma$, with $\gamma$ denoting the missing
bond component, the flux operator is
\begin{equation}
\begin{aligned}
\widehat W_p
&=\prod_{\langle rr'\rangle_\alpha\in\partial p}
  \sigma_r^\alpha\sigma_{r'}^\alpha\\
&=-\prod_{r\in p}\sigma_r^\gamma,
\end{aligned}
\label{eq:Wp}
\end{equation}
with eigenvalues $W_p=-1$ and $+1$ corresponding to $\pi$ and zero flux.
Since $[H_K,\widehat W_p]=0$, the zero-field Hilbert space separates
into flux sectors that exact imaginary-time evolution does not mix.
We therefore prepare initial states with prescribed plaquette fluxes.
At the isotropic point, the reflection symmetries of this coloring
select the homogeneous
$W_p=-1$ ground-state background~\cite{HPL2025,Lieb1994}.

\subsection{Loop-gas initialization and virtual flux insertions}
\label{subsec:lg}

\begin{figure*}[tb]
  \centering
  \IfFileExists{figures/k83_loop_gas.png}{%
    \includegraphics[width=0.9\linewidth]{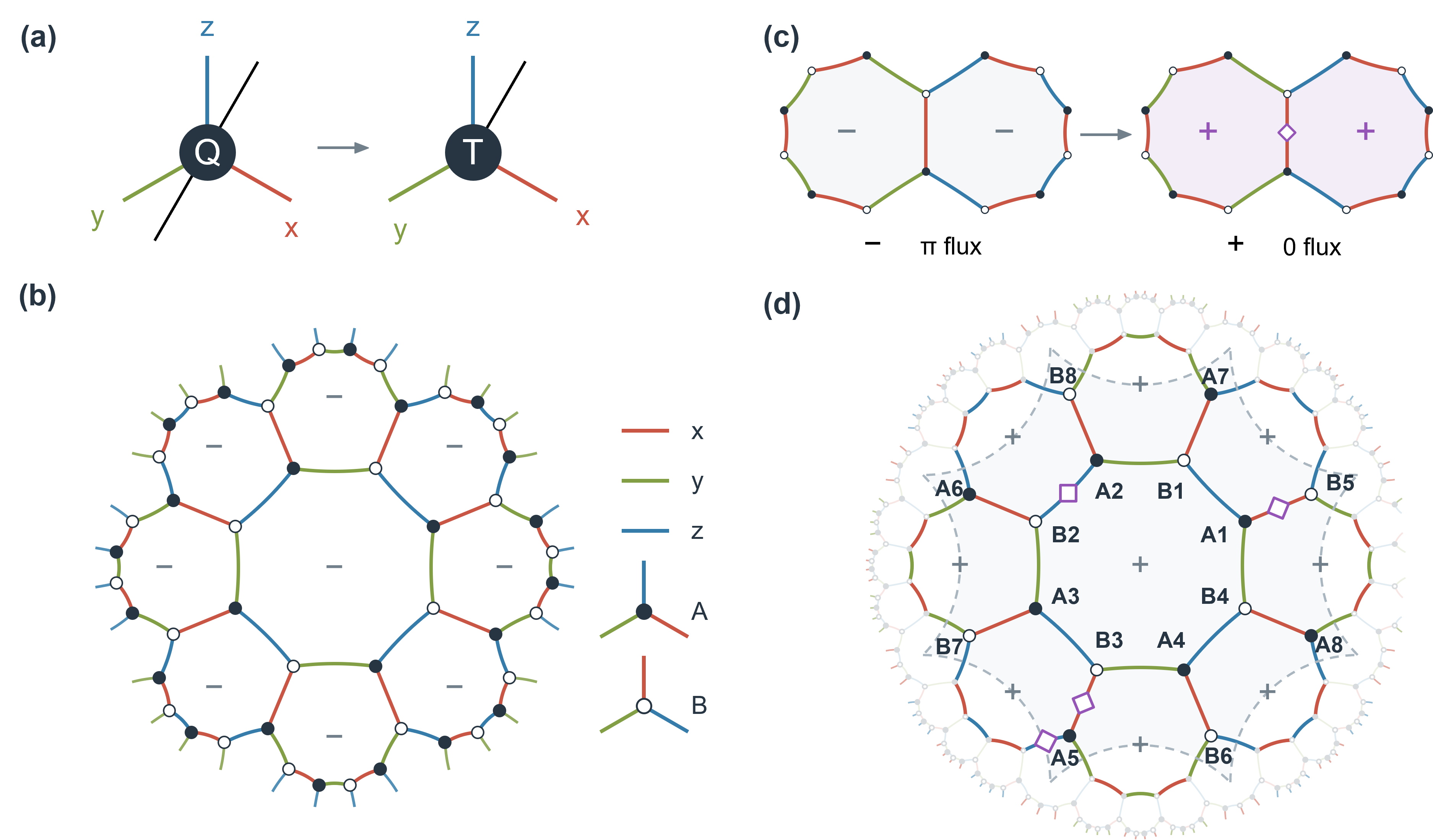}%
  }{%
    \IfFileExists{upload/Fig1(1).png}{%
      \includegraphics[width=0.9\linewidth]{upload/Fig1(1).png}%
    }{%
      \fbox{\parbox[c][42mm][c]{0.92\linewidth}{\centering\small
      Loop-gas tensors, virtual flux insertion, and the two uniform backgrounds.}}%
    }%
  }
  \caption{Loop-gas construction on the Kekul\'e-colored $\{8,3\}$ lattice.
  (a) The local LG operator $Q$ acts on a reference spin to form the PEPS
  tensor $T$. (b) Uniform $\pi$ flux, $W_p=-1$, with the two sublattices
  and bond colors indicated. (c) A virtual-bond insertion $F$ reverses
  the fluxes of the two adjacent plaquettes.
  (d) Periodic insertions in a 16-site cell generate uniform zero flux,
  $W_p=+1$. Purple symbols mark the inserted bonds.}
  \label{fig:loop_gas}
\end{figure*}

A generic product-state initialization encodes neither a prescribed
flux background nor the entanglement of the spin liquid. To supply
both before imaginary-time optimization, we construct loop-gas (LG)
initial states by applying a tensor-network operator to a uniform
reference product state~\cite{LG2019,LG2020},
\begin{equation}
\begin{aligned}
|\Psi_{\rm LG}\rangle&=\widehat Q_{\rm LG}|\Phi_0\rangle,
& |\Phi_0\rangle&=\bigotimes_r|n\rangle_r,\\
\widehat Q_{\rm LG}&=\operatorname{tTr}\prod_r Q_r.
\end{aligned}
\label{eq:LGstate}
\end{equation}
Here $\operatorname{tTr}$ contracts the virtual indices along the
lattice bonds. Locally, $Q$ acts on the reference spin $|n\rangle$
to form the PEPS tensor $T^s_{ijk}=\sum_{s'}Q^{ss'}_{ijk}n_{s'}$,
with $n_s=\langle s|n\rangle$ [Fig.~\ref{fig:loop_gas}(a)].

For the standard $(x,y,z)$ ordering, the LG tensor is
\begin{equation}
Q_{ijk}=\tau_{ijk}
(\sigma^x)^{1-i}(\sigma^y)^{1-j}(\sigma^z)^{1-k},
\label{eq:Qtensor}
\end{equation}
where $\tau_{000}=-i$ and
$\tau_{011}=\tau_{101}=\tau_{110}=1$, with all other components zero.
The binary virtual indices give an initial bond dimension $D=2$,
and the four nonzero configurations obey an even-parity constraint,
producing the loop structure. With the virtual-leg ordering and phases
on the two sublattices adapted to the local Kekul\'e coloring, repeating
the $A$ and $B$ tensors yields the uniform $W_p=-1$ state shown in
Fig.~\ref{fig:loop_gas}(b).

Other flux backgrounds can be generated through the virtual degrees
of freedom. Inserting $F=\operatorname{diag}(-1,1)$ on a bond of type
$\gamma$ is equivalent to the one-sided modification
$Q'_{ijk}=(-1)^{1-i_\gamma}Q_{ijk}$, where $i_\gamma$ is the virtual
index associated with that bond. A single insertion reverses the
fluxes of the two adjacent plaquettes [Fig.~\ref{fig:loop_gas}(c)].
For a set of insertions, the resulting flux is therefore
\begin{equation}
W'_p=(-1)^{N_p}W_p,
\label{eq:fluxparity}
\end{equation}
where $N_p$ counts the inserted bonds along the plaquette boundary.

Periodic insertions must respect the noncommuting translations of
the hyperbolic lattice. We use the 16-site cell compatible with the
Kekul\'e coloring of Ref.~\cite{HPL2025} and extend the insertions
by its hyperbolic translations. In Fig.~\ref{fig:loop_gas}(d), each
of the six inequivalent plaquettes has odd insertion parity,
converting the uniform $W_p=-1$ background into uniform $W_p=+1$.
Thus, the same local LG construction yields initial PEPS with two
distinct uniform flux backgrounds, allowing their optimized energies
to be compared in Sec.~\ref{subsec:isotropic}. Appendix~\ref{app:lg}
gives the local tensor identities and the plaquette-by-plaquette
insertion check.

\section{Tensor-network optimization and contraction}
\label{sec:method}

The iPEPS representation targets the hyperbolic bulk without relying on the
boundary-dominated averages of finite open clusters. Contracting its
norm network, however, requires an environment adapted to the
three-coordinated $\{8,3\}$ lattice and its octagonal loops.
Corner transfer matrices and their renormalization-group extensions
were originally developed for regular lattices
~\cite{Baxter1981,NishinoOkunishi1996,OrusVidal2009}.
Related contractions have been developed for non-square and hyperbolic
lattices~\cite{UedaEtAl2007,LukinSotnikov2023,LukinSotnikov2024,YangCorboz2026}.
Here we develop a geometry-adapted CTMRG (GA-CTMRG) directly on the
native quantum iPEPS tensor graph, without first regrouping sites
into an effective four-coordinated network. The key is to identify
boundary regions that retain their form under local absorption and
truncation. For $\{8,3\}$, a closed recursion involves only corner
($C$), edge ($E$), and auxiliary-corner ($aC$) environments
[Fig.~\ref{fig:hctmrg}]. We first outline the imaginary-time
optimization and then describe how this recursion yields bulk
observables.

\subsection{Imaginary-time evolution and PEPS optimization}
\label{subsec:imaginary_time}

Starting from the LG states of Sec.~\ref{sec:model_states}, we optimize
the iPEPS by imaginary-time evolution under $H_K$ at zero field and the
full Zeeman Hamiltonian at finite field. Local imaginary-time gates update the PEPS
tensors, with enlarged virtual bonds truncated back to dimension $D$~\cite{JiangWengXiang2008}.
For the reported zero-field comparison, the $\pi$-flux calculations use
simple updates, while the zero-flux calculations also include a
variational refinement.
GA-CTMRG supplies the bulk environments used to evaluate
observables of the optimized states.

\subsection{Geometry-adapted CTMRG environments and updates}
\label{subsec:contraction}

\begin{figure*}[tbp]
  \centering
  \includegraphics[width=0.9\linewidth]{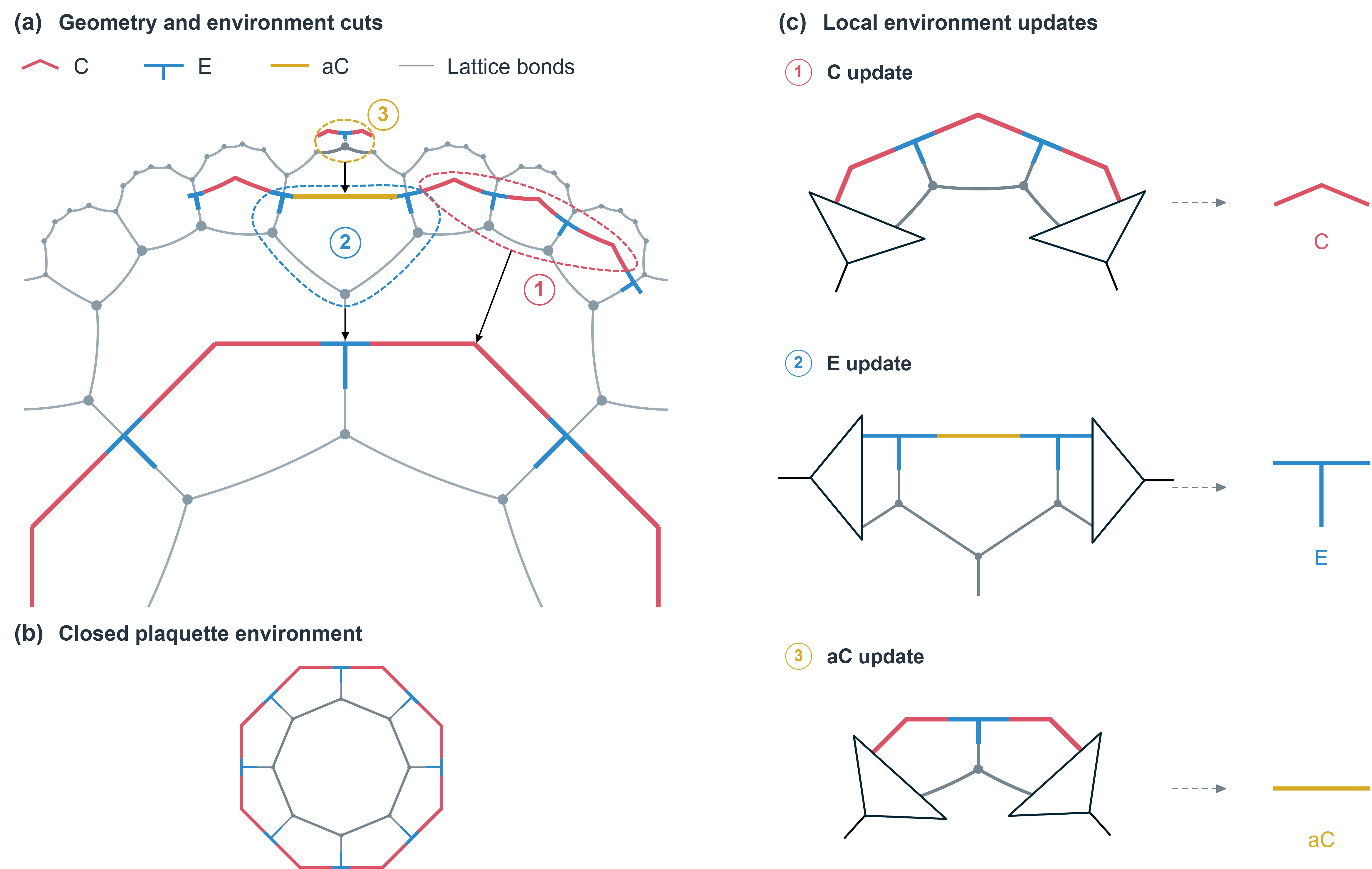}
  \caption{Geometry-adapted CTMRG (GA-CTMRG) on the native $\{8,3\}$
  network. (a) The cuts defining $C$ (pink), $E$ (blue), and $aC$
  (yellow), and the local absorption regions. Gray vertices represent
  double-layer tensors. (b) The closed environment of a central
  octagonal plaquette used to construct the boundary matrices $L$ and
  $R$. (c) Local environment updates: absorption preserves each tensor
  type, and the paired maps at the triangles compress the enlarged
  boundary indices back to $\chi$. The PEPS bond dimension is $D$.}
  \label{fig:hctmrg}
\end{figure*}

Contracting the physical indices of the PEPS tensor and its
conjugate forms a double-layer network with virtual legs of dimension
$D^2$. This network retains the original three-coordinated connectivity
and octagonal loops [gray vertices in Fig.~\ref{fig:hctmrg}(a)].
This allows its environment to be constructed from cuts of the
original tiling.

The cuts in Fig.~\ref{fig:hctmrg}(a) define three types of contracted
regions. The corner $C$ represents a region bounded by two renormalized
cuts. The edge $E$ represents a boundary strip with one double-layer
leg left open to the central network. The auxiliary corner $aC$
completes the region between neighboring edge pieces needed to advance
this boundary. With environment dimension $\chi$, $C$ and $aC$ are
$\chi\times\chi$ matrices, whereas $E$ has dimensions
$\chi\times\chi\times D^2$.
Labels for inequivalent sites and orientations are suppressed.

These three region types close under the local absorptions in
Fig.~\ref{fig:hctmrg}(c). A $C$ update absorbs two neighboring
double-layer tensors together with the surrounding $C$ and $E$
environments. An $E$ update absorbs a three-site region bounded by
two $E$ tensors and one $aC$, while an $aC$ update absorbs a one-site
region bounded by two $C$ tensors and one $E$.
In each case, absorption enlarges the two boundary indices while
preserving the pattern of exposed lattice legs of the target
environment. Compressing these indices back to $\chi$ therefore
restores the same tensor type, closing the recursion on $C$, $E$,
and $aC$.

This closure makes the contraction economical. Each update combines
regions already represented by renormalized environments, so the
number of lattice sites encoded in these regions increases rapidly
with the absorption depth. On a hyperbolic tiling, the number of sites
within successive geometric layers grows exponentially, yet the
recursion retains a finite set of environment tensors with boundary
dimensions limited by $\chi$. At fixed $D$ and $\chi$, the work of an
update is therefore governed by local tensor contractions and the
number of inequivalent environments, rather than by explicitly
contracting an exponentially growing open cluster. The repeated
absorption and compression constitute a real-space renormalization
of the bulk environment. The economy of this construction is also
reflected in the modest environment dimensions needed for the
quantum-state calculations, as discussed below.

The compression is determined by contracting the closed plaquette
environment in Fig.~\ref{fig:hctmrg}(b).
Each bond of this effective octagonal loop is a composite boundary
bond containing double-layer and environment degrees of freedom.
For a chosen target bond, we open the octagonal environment loop
at the opposite bond and contract the resulting two halves
toward the target to obtain the boundary matrices $L$ and $R$.
This partitioning of a closed tensor loop is related to
loop-based approaches such as Loop-TNR~\cite{YangGuWen2017};
here it serves to construct the CTMRG boundary-compression maps.
We then form $LR$ and retain its largest $\chi$ singular values,
\begin{equation}
LR\simeq U_\chi S_\chi V_\chi^\dagger.
\label{eq:svdLR}
\end{equation}
Following the usual singular-value truncation strategy~\cite{CorbozRiceTroyer2014}, the paired boundary-compression maps are
\begin{equation}
\begin{aligned}
P_{\rm env}&=R V_\chi S_\chi^{-1/2},\\
Q_{\rm env}&=S_\chi^{-1/2}U_\chi^\dagger L.
\end{aligned}
\label{eq:PQ}
\end{equation}
For nonzero retained singular values, the paired maps satisfy
$Q_{\rm env}P_{\rm env}=I_\chi$ on the retained subspace.
Inserting the appropriately oriented maps at the triangles in
Fig.~\ref{fig:hctmrg}(c) returns the enlarged boundary indices to
$\chi$. The retained boundary space is thus selected by the
contraction of the surrounding plaquette environment.

All three new environments are computed from the previous iteration
before they are replaced simultaneously. After rescaling the updated
tensors for numerical stability, we repeat absorption and
environment-determined compression until the rescaled environments,
retained singular-value spectra, and local observables stabilize.
This self-consistent iteration updates $C$, $E$, and $aC$; the PEPS
tensors remain fixed during the contraction.

\subsection{Measurements and numerical control}
\label{subsec:measurements}

The converged environments close the central plaquette network in
Fig.~\ref{fig:hctmrg}(b). Local observables are evaluated by inserting
the corresponding operators and normalizing by the uninserted network
contraction.
Bond insertions give spin correlators and, through Eq.~\eqref{eq:HK},
the exchange energy per site. The plaquette flux is evaluated by
inserting the eight-site operator in Eq.~\eqref{eq:Wp} across the
central octagon. Averages include the appropriate multiplicities of
inequivalent bonds and plaquettes in the tensor cell.

The PEPS bond dimension $D$ and environment dimension $\chi$
are the principal truncation parameters. For the bond dimensions
studied, local observables are stable across several $\chi$ values,
including $\chi=2D$ and $2D+10$. Thus, modest environment
dimensions proportional to $D$ suffice in these calculations,
despite the $D^2$ bond dimension of the double-layer network.
Implementation and cell-average conventions are given in
Appendix~\ref{app:numerics}. Classical and quantum Ising tests in
Appendix~\ref{app:ising} further support using the same
environment recursion to evaluate magnetic observables on the
native lattice.

\section{Zero-field hyperbolic Kitaev spin liquids}
\label{sec:zero_field}

At zero field, conserved plaquette fluxes provide a well-defined
reference for the variational states. This lets us distinguish the
energetics of competing homogeneous-flux initializations from the
response of bond correlations within a prescribed background.

\subsection{Isotropic ground state and flux-sector comparison}
\label{subsec:isotropic}

\begin{figure}[htbp]
  \centering
  \includegraphics[width=\linewidth]{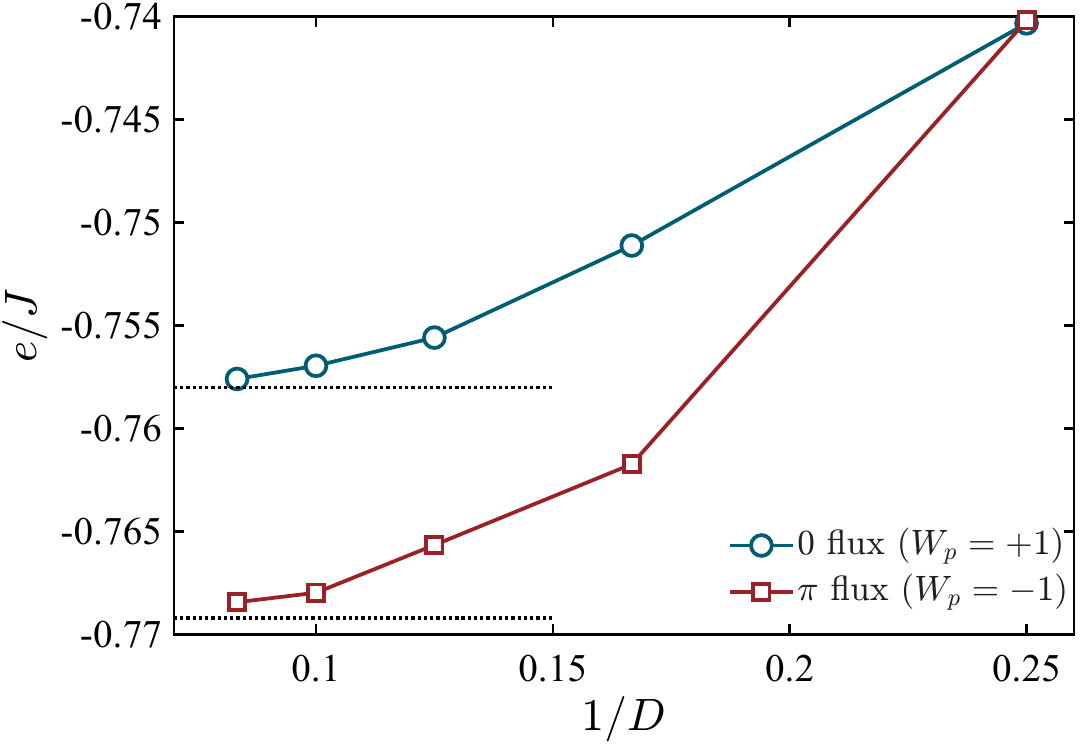}
  \caption{Optimized energy per site versus inverse PEPS bond dimension at
  $J_x=J_y=J_z=J$ and $h=0$. Circles denote the zero-flux
  initialization ($W_p=+1$), and squares denote the $\pi$-flux
  initialization ($W_p=-1$). Both branches include $D=4,6,8,10,12$.
  Solid lines connect the sampled values. Dotted horizontal lines
  indicate the reference energy levels for the two backgrounds
  reported in Ref.~\cite{HPL2025}.}
  \label{fig:k83_zero_field_energy}
\end{figure}

For the isotropic model with $J_x=J_y=J_z=J>0$, we express the energy
per site in units of $J$.
Figure~\ref{fig:k83_zero_field_energy} compares the energies $e_-$ and $e_+$
obtained from uniform $\pi$-flux ($W_p=-1$) and zero-flux ($W_p=+1$)
initializations, respectively, as a function of $1/D$.

The $\pi$-flux branch is energetically favored at $D=6,8,10,12$,
while the zero-flux branch is marginally lower at $D=4$.
At $D=10$, we obtain $e_-/J\simeq-0.767985$ and
$e_+-e_-\simeq1.1\times10^{-2}J$ per site.
The ordering at the larger bond dimensions agrees with the established uniform
$\pi$-flux ground state~\cite{HPL2025}.

\subsection{Anisotropy along the \texorpdfstring{$J_x=J_y$}{Jx=Jy} line}
\label{subsec:anisotropy}

The exactly solved $\{8,3\}$ Kitaev model exhibits a Majorana
metal at isotropic exchange and a gapped Abelian $\mathbb{Z}_2$
spin liquid at sufficiently strong anisotropy, with a hyperbolic
toric-code description in the strongly anisotropic limit~\cite{HPL2025}.
Both belong to the same $\mathbb{Z}_2$ gauge framework, but have
different Majorana matter spectra. 
Here we examine the evolution of local spin correlations
under exchange anisotropy. We vary
$r=J_z/J_x$ with $J_x=J_y$, keeping
$J_x+J_y+J_z=3J$ fixed.

\begin{figure}[htbp]
  \centering
  \includegraphics[width=\linewidth]{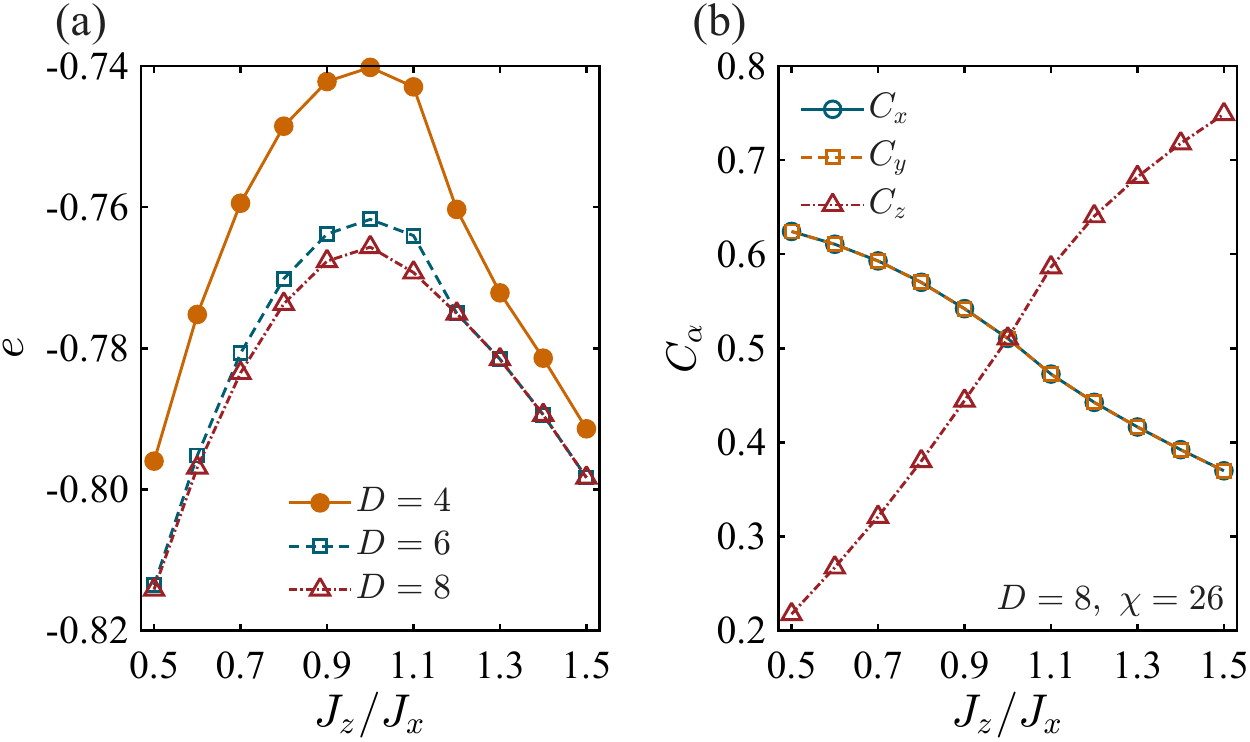}
  \caption{Zero-field exchange-anisotropy scan on the Kekul\'e-colored
  $\{8,3\}$ lattice with $J_x=J_y$ and fixed
  $J_x+J_y+J_z=3J$.
  (a) Optimized energy per site $e$ versus $J_z/J_x$ for PEPS bond
  dimensions $D=4,6,8$.
  (b) Nearest-neighbor correlations $C_\alpha={\langle\sigma_i^\alpha\sigma_j^\alpha\rangle}_{\alpha\mathrm{-bond}}$, averaged over
  $\alpha$-type bonds, at $D=8$ and environment dimension $\chi=26$.
  The plaquette flux remains $\langle\widehat W_p\rangle=-1$ along
  the entire scan.}
  \label{fig:k83_anisotropy}
\end{figure}

Figure~\ref{fig:k83_anisotropy}(a) shows the optimized energies
for $D=4,6,8$, with the $D=6$ and $D=8$ results nearly coincident.
At isotropy, the three bond correlations are equal. As $r$ increases,
$C_z$ grows while $C_x=C_y$ decreases
[Fig.~\ref{fig:k83_anisotropy}(b)], reflecting the redistribution
of short-range spin correlations toward the stronger $z$ bonds.
With the total exchange scale held fixed, the equality of bond
correlations at isotropy is also consistent with the nearly
stationary energy around $r=1$ in panel (a). Throughout the scan,
$\langle\widehat W_p\rangle=-1$: the changing local correlations
occur within an unchanged uniform $\pi$-flux background.
This is consistent with the exactly solved picture in which the
Majorana metal and anisotropic gapped $\mathbb{Z}_2$ spin liquid
can share the same flux sector. 

The isotropic $\pi$-flux state then provides the reference for the
real-field response, where plaquette flux is no longer conserved.

\section{Response to uniform and staggered [111] fields}
\label{sec:k83_field}

\subsection{Real Zeeman fields and measured responses}
\label{subsec:field_definition}

We now consider the isotropic ferromagnetic model in a real $[111]$
Zeeman field,
\begin{equation}
H(h)=H_K-h\sum_r\eta_r
(\sigma_r^x+\sigma_r^y+\sigma_r^z),
\label{eq:k83_zeeman}
\end{equation}
where $\eta_r=1$ for the uniform field and $\eta_r=s_r$ for the
staggered field, with $s_A=+1$ and $s_B=-1$ on the two sublattices.
The local Zeeman field is $\boldsymbol{h}_r=\eta_r h(1,1,1)$,
with magnitude $\sqrt{3}h$. We measure the energy per site
$e=\langle H(h)\rangle/N$,
the cell-averaged plaquette flux
$\overline W=\frac{1}{6}\sum_{p=1}^{6}\langle\widehat W_p\rangle$,
and the polarization along the local field,
\begin{equation}
P=\frac{1}{N}\sum_r\eta_r
\left\langle\frac{\sigma_r^x+\sigma_r^y+\sigma_r^z}{\sqrt{3}}\right\rangle.
\label{eq:k83_field_polarization}
\end{equation}
The Zeeman term mixes flux sectors, allowing $\overline W$ to take
intermediate values at finite $h$.

For the Kekul\'e-colored honeycomb model, field-induced three-spin
terms have been derived for both uniform and staggered
fields~\cite{NegahdariLangari2025}; staggered-field effects on the
ordinary honeycomb model have also been studied~\cite{NakazawaKatoMotome2022}.
The two field patterns also impose different symmetry constraints.
A color-preserving reflection $\mathcal P$ of the Kekul\'e lattice
exchanges the two sublattices and reverses plaquette orientation.
The uniform-field Hamiltonian preserves $\mathcal P$, whereas the
staggered-field Hamiltonian preserves the combined symmetry
$\mathcal P\Theta$, where $\Theta$ denotes spin time reversal.
This distinction constrains possible chiral responses, as discussed
below and in Appendix~\ref{app:third_order}.

\subsection{Uniform-field response}
\label{subsec:uniform}

\begin{figure*}[tbp]
  \centering
  \includegraphics[width=\linewidth]{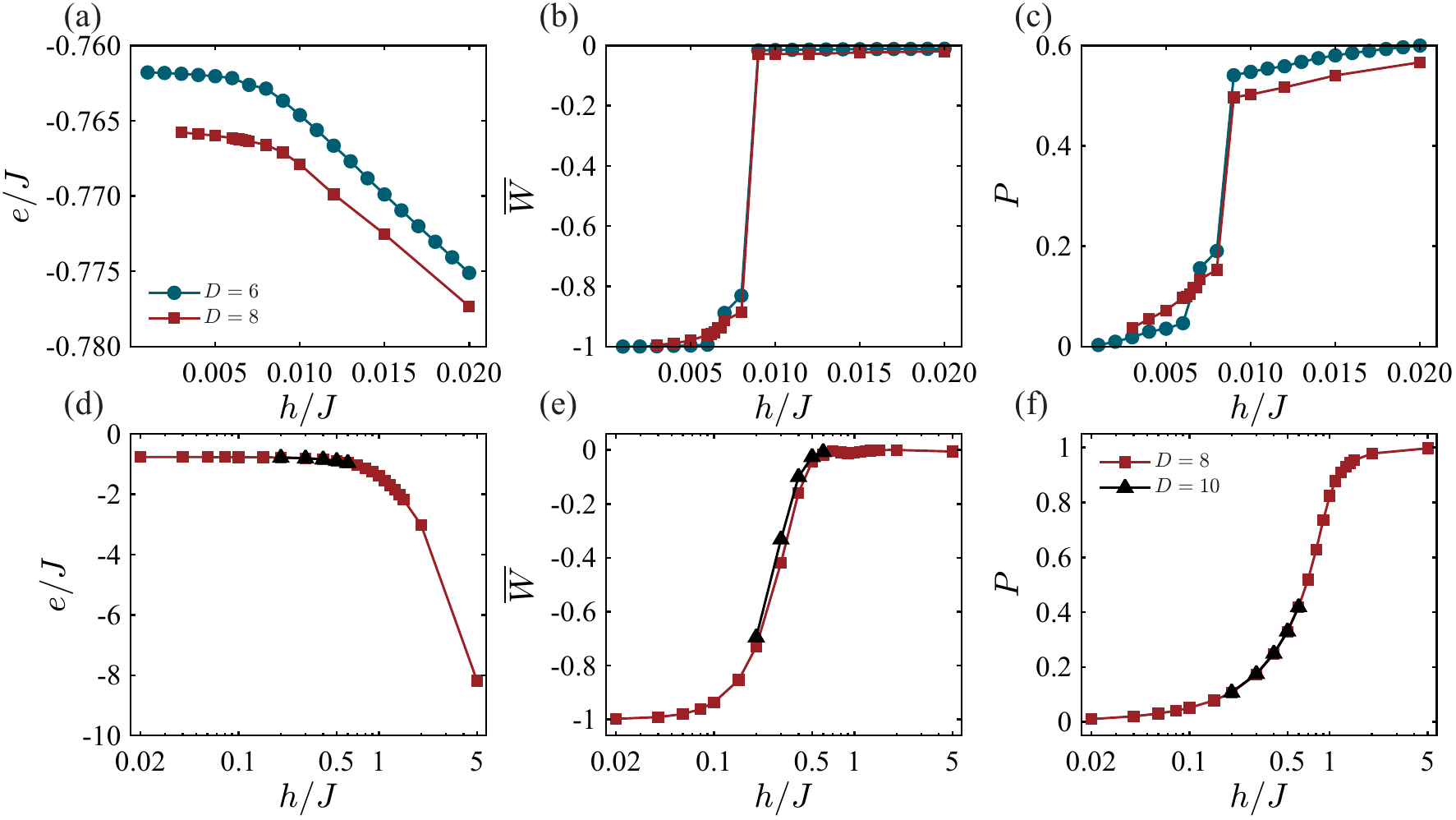}
  \caption{iPEPS response to uniform and staggered $[111]$ Zeeman fields.
  (a)--(c) Uniform-field energy per site, mean plaquette flux, and
  field-aligned polarization at $D=6$ (circles) and $D=8$ (squares).
  For both bond dimensions, the points at $h/J=0.008$ and $0.009$
  come from the low-field continuation and the more strongly
  polarized solution, respectively, selected by energy from the
  two initializations.
  (d)--(f) The ascending staggered-field sweep at $D=8$, with $D=10$
  results (triangles) over the overlapping range. The bottom row uses
  a logarithmic field axis. Lines connect the sampled points.}
  \label{fig:k83_field_tn}
\end{figure*}

The uniform-field PEPS solutions change sharply between $h/J=0.008$
and $0.009$ at both $D=6$ and $8$
[Fig.~\ref{fig:k83_field_tn}(a)--(c)]. Across these sampled fields,
the plotted polarization rises from $P\simeq0.15$--$0.19$ to
$P\simeq0.50$--$0.54$, and the mean plaquette flux approaches zero.
The plotted points on the low-field side come from states continued
from zero field, while those on the high-field side come from a
different initialization that yields more strongly polarized states.
The energetically preferred variational branch switches
across this interval, so the apparent jumps in $P$ and $\overline W$
reflect the same selection of variational solutions (see
Appendix~\ref{app:uniform_initialization}).

Away from a transition, the Hellmann--Feynman relation
$de_0/dh=-\sqrt{3}\,P_0$ connects the ground-state energy slope to
the ground-state polarization. A finite discontinuity in $P_0$ would
give a kink in $e_0(h)$ and
identify a first-order transition. A first-order interpretation of
the uniform-field response therefore remains conditional on
establishing a ground-state polarization jump independently of the
selection between initializations.

\subsection{Staggered-field evolution}
\label{subsec:staggered}

On the ascending sweep, the energy, mean plaquette flux, and
polarization evolve continuously within the numerical resolution of
the sampled points in the region of rapid change
[Fig.~\ref{fig:k83_field_tn}(d)--(f)]. Such smooth evolution does not
by itself distinguish a continuous transition from a crossover;
an independent signature of criticality would be required.

The field scale of this response is substantially larger than in the
uniform case. For $D=8$, the mean flux rises from near $-1$ at weak field
to approximately $-0.42$ at $h/J=0.3$ and $-0.06$ at $h/J=0.5$.
The polarization continues to increase toward saturation at stronger
fields. The available $D=10$ results follow these trends over the
overlapping field range.


\subsection{Exchange-energy balance and weak-field interpretation}
\label{subsec:field_interpretation}

The different field scales have a simple energetic component.
In both cases, the energy per site separates into exchange and
Zeeman contributions, $e=e_K-\sqrt{3}\,hP$, where
$e_K=\langle H_K\rangle/N$. For product states fully polarized
along their local fields, the exchange energies per site are $-J/2$
for the uniform pattern and $+J/2$ for the staggered pattern, while
the Zeeman contribution is $-\sqrt{3}h$ in both cases.
Thus staggered polarization incurs an exchange cost relative to
uniform polarization.
Although these product-state values do not predict a transition field,
they help explain why the staggered response occurs on a substantially
larger field scale.

Exchange energetics alone do not describe how the fields perturb the
fractionalized excitations of the zero-field Majorana metal, whose
low-energy density of states (DOS) is dominated by non-Abelian
hyperbolic Bloch states~\cite{HPL2025}. To examine this second aspect,
we project the third-order Zeeman perturbation into the uniform
$\pi$-flux sector, retain its quadratic Majorana part, and approximate
the virtual-sector energy denominators by a common value.
For an oriented two-bond path $i\xrightarrow{\alpha}j
\xrightarrow{\beta}k$, the Kekul\'e bond-color ordering gives
$q_j=\varepsilon_{\alpha\beta\gamma}=s_j$ in the chosen plaquette
orientation, where $\gamma$ is the third bond color at the central site.
The three local Zeeman signs multiply to $1$ for the uniform field,
but to $s_i s_j s_k=s_j$ for the staggered
field. In the same fixed-flux bond gauge, these sign structures yield
distinct single-particle perturbations,
\begin{equation}
\delta\mathcal H_{\rm u}^{(3)}=-\kappa\Gamma\mathsf W,
\qquad
\delta\mathcal H_{\rm s}^{(3)}=\kappa\mathsf W.
\label{eq:k83_effective_hopping}
\end{equation}
Here $\mathsf W$ is the dimensionless single-particle matrix multiplying
the oriented $K$ term of Ref.~\cite{HPL2025},
$\Gamma=\operatorname{diag}(s_r)$ acts on the hopping endpoints,
and $\kappa$ is the effective third-order hopping scale
(Appendix~\ref{app:third_order}). A sublattice-alternating sign
pattern of next-nearest-neighbor Majorana hoppings also occurs in a
staggered three-spin model on the ordinary honeycomb
lattice~\cite{SoaresPereira2026}. The reflection symmetry discussed
above also distinguishes their possible chiral responses: a state
preserving $\mathcal P$ has vanishing net reflection-odd chirality,
whereas $\mathcal P\Theta$ symmetry permits such a response.
Neither symmetry statement alone identifies a chiral phase.

\begin{figure}[htbp]
  \centering
  \includegraphics[width=\linewidth]{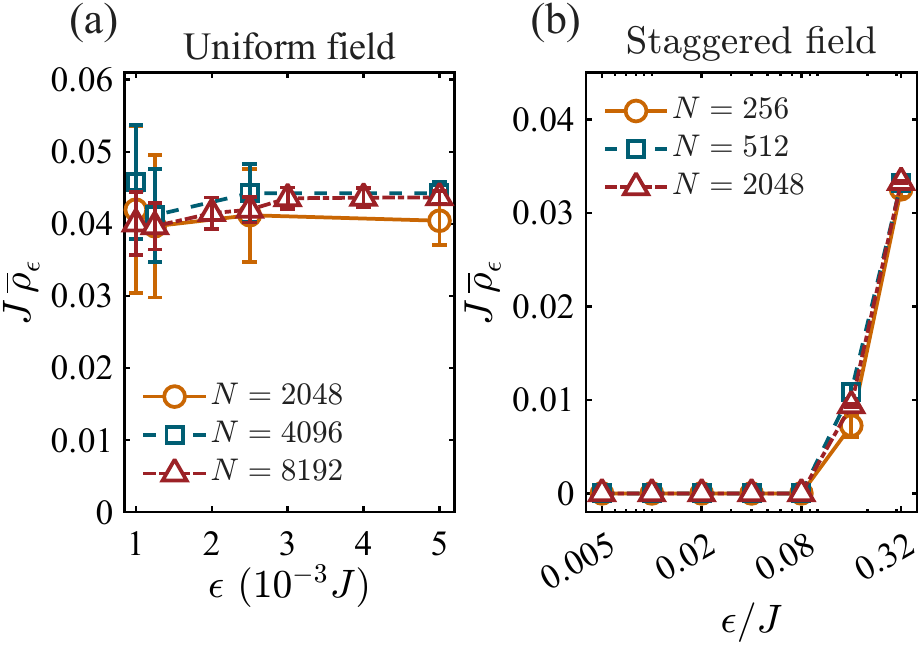}
  \caption{Window-averaged Majorana DOS in the fixed $\pi$-flux
  effective models at $\kappa/J=0.03$.
  (a) The uniform-type hopping $-\kappa\Gamma\mathsf W$, for
  $N=2048,4096,8192$.
  (b) The staggered-type hopping $\kappa\mathsf W$, for $N=256,512,2048$.
  The plotted quantity is $J\overline\rho_\epsilon$, with
  $\overline\rho_\epsilon=N_{|E|<\epsilon}/(2\epsilon N)$ averaged
  over boundary twists. Error bars are standard errors over
  independent twist pairs. The two panels use different window ranges
  and linear/logarithmic horizontal axes, respectively.
  Sampling details are given in Appendix~\ref{app:dos_sampling}.}
  \label{fig:k83_field_dos}
\end{figure}

The spectral distinction is visible in the sampled effective models.
At $\kappa/J=0.03$, the uniform-type hopping retains a window-averaged
DOS of approximately $0.04/J$ down to $\epsilon/J=10^{-3}$
[Fig.~\ref{fig:k83_field_dos}(a)]. The staggered-type hopping has no
sampled states in the windows $\epsilon/J\leq0.08$ shown in
Fig.~\ref{fig:k83_field_dos}(b), with weight appearing at larger windows.
This depletion is consistent with the gapped solvable $K$ model
of Ref.~\cite{HPL2025}. Since these spectra come from a fixed-flux
quadratic approximation, they do not determine the excitation gap or
topological order of the full spin Hamiltonian in a real Zeeman field.

\section{Discussion and conclusions}
\label{sec:conclusions}

Uniform and staggered $[111]$ fields of equal local magnitude produce
markedly different responses from the same isotropic hyperbolic Kitaev
spin-liquid reference state. Under the uniform field, the energy-selected
variational solution changes sharply at weak field, whereas the
staggered-field polarization and mean plaquette flux evolve over a much
broader field range. At zero field, the Kekul\'e-colored
$\{8,3\}$ model hosts a Majorana metal in the uniform
$\pi$-flux background~\cite{HPL2025}. A physical Zeeman field, however, mixes flux
sectors, so its many-body response cannot be inferred from a
single fixed-flux Majorana spectrum alone.

Two complementary aspects help explain this field-pattern dependence.
Ferromagnetic exchange favors uniform over staggered polarization,
providing a qualitative explanation for the separation of their
response scales. Beyond this energetic competition, the field pattern
combines with the Kekul\'e bond coloring to produce different
third-order Majorana hoppings and low-energy effective spectra.
The different symmetry constraints further restrict possible
chiral responses without determining the field-induced phases. Thus the
spatial structure of the perturbation matters both for polarization
energetics and for its coupling to fractionalized degrees of freedom.

The fixed-$\pi$-flux effective models reveal how the two field
patterns modify Majorana motion, but do not capture the flux
fluctuations induced by a physical Zeeman field. Their contrasting
spectra therefore do not establish the excitation gap or
topological order of the full spin system. Likewise, the
uniform- and staggered-field responses do not yet determine
the nature of the possible phase transitions.
By combining flux-resolved iPEPS with geometry-adapted bulk
contraction, we provide an economical numerical framework
for studying correlated quantum spins on hyperbolic lattices
without relying on large open clusters. Beyond this
methodological advance, our results connect the
geometry-dependent Majorana and flux structure of an exactly
solvable quantum spin liquid to its many-body response under
physical perturbations that break flux conservation.
This provides a route to understanding how fractionalized
quantum matter evolves beyond its exactly solvable limits.

\noindent\textit{Note added.} We recently became aware of independent work by
Lukin and Sotnikov~\cite{LukinSotnikov2026}, which formulates CTMRG
on general planar graphs using local message-passing equations
and corner matrices, with classical Ising benchmarks including
the hyperbolic $\{8,3\}$ lattice. Our independently developed
GA-CTMRG uses a closed recursion of corner ($C$), edge ($E$),
and auxiliary-corner ($aC$) environments on the native
$\{8,3\}$ geometry, with boundary compression determined from
an octagonal plaquette environment. We apply this construction
to flux-resolved quantum iPEPS and the Kitaev model under
physical magnetic fields.

\begin{acknowledgments}
This work was supported by the National Key R\&D Program of China
(Grants No.~2022YFA1403601, No.~2023YFA1406500,
No.~2024YFA1408604, No.~2024YFA1410500),
the Scientific Research Innovation Capability Support Project for
Young Faculty (Grant No.~SRICSPYF-ZY2025164),
the National Natural Science Foundation of China
(Grants No.~12322402, No.~12274206, No.~12274458),
the Quantum Science and Technology National Science and Technology
Major Project (Grant No.~2021ZD0302800),
the Natural Science Foundation of Jiangsu Province
(Grant No.~BK20233001), and the Fundamental Research Funds
for the Central Universities (Grant No.~KG202501).
\end{acknowledgments}

\appendix
\renewcommand{\theequation}{\Alph{section}\arabic{equation}}

\section{Classical and quantum Ising benchmarks}
\label{app:ising}

\subsection{Classical Ising model}
\label{app:classical_ising}

\begin{figure}[htbp]
  \centering
  \includegraphics[width=\linewidth]{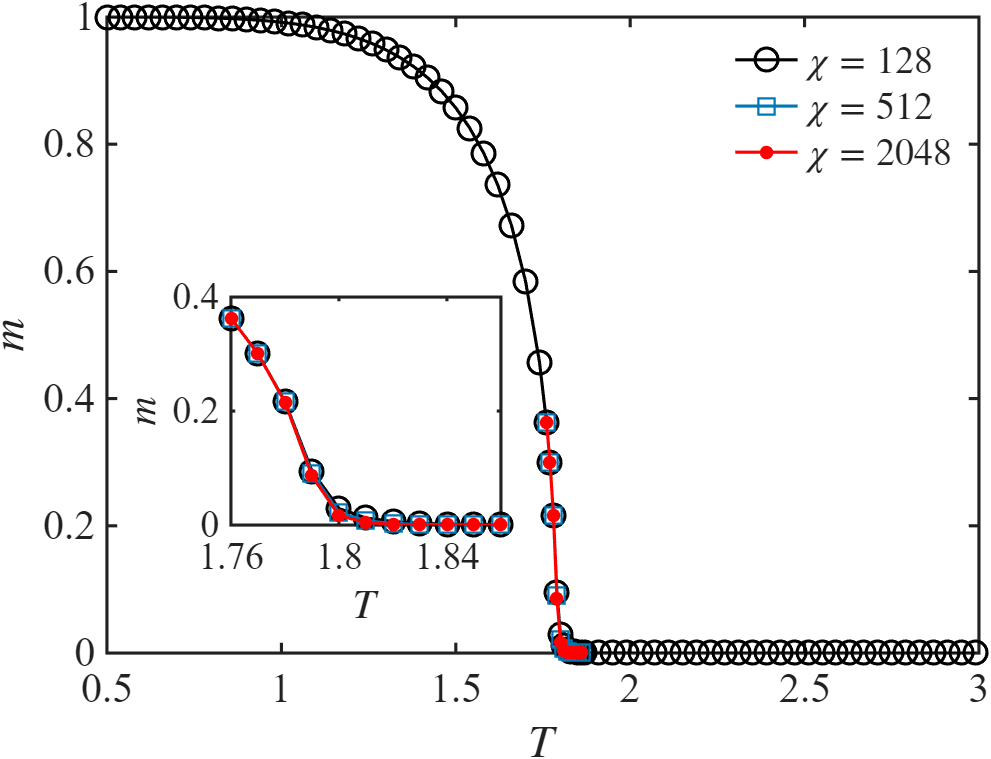}
  \caption{Bulk magnetization of the classical ferromagnetic Ising
  model on the hyperbolic $\{8,3\}$ lattice as a function of
  temperature $T/J_{\rm I}$ (with $J_{\rm I}=1$). Data for
  environment dimensions $\chi=128$
  (open circles), $512$ (open squares), and $2048$ (filled circles)
  are compared. The inset enlarges the region of the rapid drop
  in magnetization. Lines guide the eye.}
  \label{fig:ising_classical}
\end{figure}

As a benchmark of the bulk contraction, we consider the
ferromagnetic classical Ising model on the $\{8,3\}$ lattice,
\begin{equation}
H_{\rm cl}=-J_{\rm I}\sum_{\langle rr'\rangle}
\varsigma_r\varsigma_{r'},\qquad \varsigma_r=\pm1,
\label{eq:ising_classical_hamiltonian}
\end{equation}
where $J_{\rm I}>0$. We set $J_{\rm I}=k_B=1$ in the
calculations, so the plotted temperature is in units of $J_{\rm I}$.
The bulk magnetization $m_{\rm cl}=|\langle\varsigma\rangle|$
is evaluated using the geometry-adapted environment recursion. Figure~\ref{fig:ising_classical}
shows an ordered low-temperature regime and a rapid suppression of
magnetization near $T/J_{\rm I}\simeq1.8$. The curves for
$\chi=128$, $512$, and $2048$ agree closely at the sampled
points, including in the transition region shown in the inset.

For comparison at the same coordination number $z=3$, the
ferromagnetic Ising model has the exact critical temperatures
$T_c^{\rm hc}/J_{\rm I}=2/\ln(2+\sqrt{3})\simeq1.519$
on the honeycomb lattice and
$T_c^{\rm Bethe}/J_{\rm I}=1/\operatorname{arctanh}(1/2)
\simeq1.8205$ on the Bethe lattice. The transition region in
Fig.~\ref{fig:ising_classical} is therefore much closer to the
Bethe reference than to the honeycomb value. This comparison
provides a useful consistency check on the hyperbolic bulk
calculation; the magnetization curves alone do not determine a
precise $T_c$ or its critical exponents.

\subsection{Transverse-field Ising model}
\label{app:quantum_ising}

As a second benchmark, we consider the transverse-field Ising model,
which has also been studied on hyperbolic lattices
~\cite{DaniskaGendiar2016}, with Hamiltonian
\begin{equation}
H_{\rm TFI}=-J_x\sum_{\langle rr'\rangle}\sigma_r^x\sigma_{r'}^x
-h_z\sum_r\sigma_r^z,
\label{eq:ising_quantum_hamiltonian}
\end{equation}
where $J_x>0$ and the spin operators are Pauli matrices.
Energies are measured in units of $J_x$, and the field is quoted as
$h_z/J_x$. We measure the ordered magnetization
$m_x=|\langle\sigma^x\rangle|$ and the transverse polarization
$p_z=|\langle\sigma^z\rangle|$ in the optimized iPEPS,
using GA-CTMRG to evaluate the corresponding spin insertions.

\begin{figure}[htbp]
  \centering
  \includegraphics[width=\linewidth]{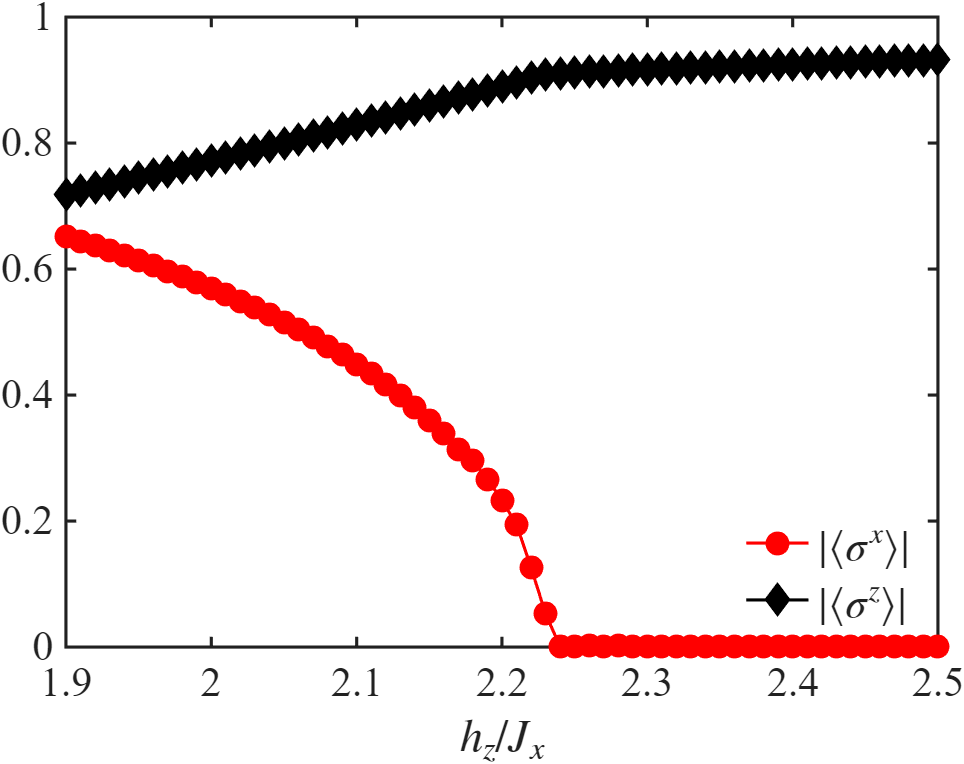}
  \caption{Transverse-field Ising response on the $\{8,3\}$ lattice.
  The ordered magnetization $m_x=|\langle\sigma^x\rangle|$ (red circles)
  decreases to zero as $h_z/J_x$ increases, while the transverse
  polarization $p_z=|\langle\sigma^z\rangle|$ (black diamonds) grows.
  Lines connect the sampled fields.}
  \label{fig:ising_quantum}
\end{figure}

As $h_z/J_x$ increases, $m_x$ vanishes, while $p_z$ continues to grow
beyond the loss of Ising order [Fig.~\ref{fig:ising_quantum}].
These data provide a qualitative benchmark of the field response;
we do not extract a critical field from the displayed curves.

\section{Loop-gas tensors and flux verification}
\label{app:lg}

\subsection{Local tensors and virtual-leg conventions}
\label{app:lg_local}

We label the virtual legs by bond color,
$(i,j,k)=(i_x,i_y,i_z)$, and contract equal binary indices on each
bond. The physical matrix indices $(s,s')$ denote output and input
in the $\sigma^z$ basis. For the flux verification, an explicit
color-ordered representative of the opposite local handedness on
the two sublattices is $Q^A=Q$ of Eq.~\eqref{eq:Qtensor} and
\begin{equation}
Q^B_{ijk}=\tau_{ijk}
(\sigma^z)^{1-k}(\sigma^y)^{1-j}(\sigma^x)^{1-i}.
\label{eq:lg_B_representative}
\end{equation}
Thus, the nonzero operator components in this representative are
\begin{equation}
\begin{array}{c|cccc}
(i_xi_yi_z)&000&011&101&110\\ \hline
Q^A& I_2&\sigma^x&\sigma^y&\sigma^z\\
Q^B&-I_2&\sigma^x&\sigma^y&\sigma^z
\end{array}
\label{eq:lg_AB_components}
\end{equation}
and all other components vanish. The relative sign of the $000$
components follows from
$\sigma^x\sigma^y\sigma^z=iI_2$ and
$\sigma^z\sigma^y\sigma^x=-iI_2$ with the same
$\tau_{000}$ as in Sec.~\ref{subsec:lg}.

Acting on a reference spin $|n\rangle$, the four components of
$T^A$ are $(n,\sigma^x n,\sigma^y n,\sigma^z n)$ in the column
order of Eq.~\eqref{eq:lg_AB_components}; those of $T^B$ have
$-n$ in the first column. The flux identities below do not depend
on the choice of $|n\rangle$, provided that the LG state is nonzero.

\subsection{Local flux identities and the uniform pi-flux state}
\label{app:lg_flux}

Introduce the virtual matrix
\begin{equation}
v=\begin{pmatrix}0&1\\ i&0\end{pmatrix}.
\label{eq:lg_virtual_v}
\end{equation}
A subscript on $v$ specifies the color of the leg on which it acts,
with $(v_\alpha Q)_{\boldsymbol i}
=\sum_{j_\alpha}v_{i_\alpha j_\alpha}
Q_{\boldsymbol i[i_\alpha\to j_\alpha]}$.
For each cyclic choice
$(\alpha,\beta,\gamma)=(x,y,z),(y,z,x),(z,x,y)$,
the component table gives
\begin{equation}
\begin{aligned}
\sigma^\gamma Q^A&=v_\alpha^*v_\beta Q^A,\\
\sigma^\gamma Q^B&=-v_\alpha v_\beta^* Q^B.
\end{aligned}
\label{eq:lg_local_flux_identity}
\end{equation}
Here $*$ denotes complex conjugation. For example,
$\sigma^z Q^A_{000}=Q^A_{110}$ and
$\sigma^z Q^A_{011}=iQ^A_{101}$; the other entries follow by the
same Pauli algebra. Multiplying by the reference spin on the right
also gives these identities for $T^A$ and $T^B$.

On a plaquette of missing color $\gamma$, the virtual matrices act
only on its alternating $\alpha$ and $\beta$ boundary legs.
At the two ends of every boundary bond, the matrices cancel via
$(v^*)^Tv=v^Tv^*=I_2$, with the transposes set by the shared
virtual-index contraction. Four sites on each octagon belong to the
$B$ sublattice, so their minus signs in
Eq.~\eqref{eq:lg_local_flux_identity} also cancel. Hence
\begin{equation}
\begin{aligned}
\left(\prod_{r\in p}\sigma_r^\gamma\right)
\widehat Q_{\rm LG}&=\widehat Q_{\rm LG},\\
\widehat W_p|\Psi_{\rm LG}\rangle&=-|\Psi_{\rm LG}\rangle.
\end{aligned}
\label{eq:lg_uniform_pi_verification}
\end{equation}
The last sign uses Eq.~\eqref{eq:Wp}, whose boundary product is
ordered consecutively around the octagon. For alternating colors,
the on-site Pauli phases multiply to
$(i\varepsilon_{\alpha\beta\gamma})^5
(-i\varepsilon_{\alpha\beta\gamma})^3=-1$.
Thus the eight-spin product has eigenvalue $+1$, while the
plaquette flux in our convention is $W_p=-1$.

\subsection{Bond insertions and the 16-site zero-flux construction}
\label{app:lg_insertions}

The virtual insertion $F=\operatorname{diag}(-1,1)$ anticommutes
with $v$ and $v^*$, giving $(v^*)^TFv=v^TFv^*=-F$ on an
inserted boundary bond. Each such bond therefore contributes one
extra minus sign under the plaquette action. If
$|\Psi_{\mathcal I}\rangle$ denotes the state with insertions
on a bond set $\mathcal I$, then
\begin{equation}
\widehat W_p|\Psi_{\mathcal I}\rangle
=-(-1)^{N_p}|\Psi_{\mathcal I}\rangle,
\label{eq:lg_inserted_flux_verification}
\end{equation}
which verifies Eq.~\eqref{eq:fluxparity} for the $\pi$-flux reference.
A single inserted edge belongs to two plaquettes and flips both
fluxes; the other plaquettes are unaffected. Since
$F_{aa}=(-1)^{1-a}$, absorbing $F$ into either endpoint gives the
one-sided tensor modification in Sec.~\ref{subsec:lg}.
Applying that modification to both endpoints instead gives
$F^2=I_2$ and removes the insertion.

For the pattern in Fig.~\ref{fig:loop_gas}(d), the four independent
inserted bonds, in the displayed site labels, are
\begin{equation}
\begin{aligned}
\mathcal I_{\rm cell}=\{& (A_1,B_5)_x,\ (A_2,B_2)_z,\\
                      & (A_5,B_5)_z,\ (A_5,B_3)_x\}.
\end{aligned}
\label{eq:lg_cell_insertions}
\end{equation}
The set is extended by the hyperbolic translations of the 16-site
cell. Table~\ref{tab:lg_cell_flux} lists the six inequivalent
plaquette walks in the quotient-site labels. Cell crossings use
the corresponding hyperbolic boundary identifications.
Counting the inserted edges on each closed walk gives
$(N_{p_1},\ldots,N_{p_6})=(1,1,3,1,1,1)$.
All six parities are odd, so
Eq.~\eqref{eq:lg_inserted_flux_verification} gives $W_p=+1$ for
every plaquette of the periodically extended state.

\begin{table}[t]
  \caption{Plaquette check for the four insertions in
  Eq.~\eqref{eq:lg_cell_insertions}. Consecutive site labels form a
  closed boundary walk, including the bond from the last site back
  to the first. The missing color is $\gamma$, and $N_p$ counts
  inserted boundary edges with their periodic identifications.
  Each row yields $W_p=-1$ before insertion and $W'_p=+1$ afterward.}
  \label{tab:lg_cell_flux}
  \begingroup
  \setlength{\tabcolsep}{3pt}
  \begin{ruledtabular}
  \begin{tabular}{cc lc}
    Plaquette & $\gamma$ & Boundary walk & $N_p$\\
    \hline
    $p_1$ & $x$ & $A_1 B_1 A_2 B_2 A_3 B_3 A_4 B_4$ & 1\\
    $p_2$ & $x$ & $A_5 B_8 A_8 B_7 A_7 B_6 A_6 B_5$ & 1\\
    $p_3$ & $y$ & $A_1 B_5 A_5 B_3 A_3 B_7 A_7 B_1$ & 3\\
    $p_4$ & $y$ & $A_2 B_8 A_8 B_4 A_4 B_6 A_6 B_2$ & 1\\
    $p_5$ & $z$ & $A_1 B_5 A_6 B_2 A_3 B_7 A_8 B_4$ & 1\\
    $p_6$ & $z$ & $A_2 B_8 A_5 B_3 A_4 B_6 A_7 B_1$ & 1
  \end{tabular}
  \end{ruledtabular}
  \endgroup
\end{table}

\section{GA-CTMRG implementation and measurements}
\label{app:numerics}

\subsection{Environment indices and local absorption}
\label{app:implementation}

For completeness, the local double-layer tensor used to contract a
fixed PEPS is
\begin{equation}
\mathcal T_{(ii')(jj')(kk')}
=\sum_s T^s_{ijk}\,\overline{T}^{s}_{i'j'k'}.
\label{eq:doubletensor}
\end{equation}
The native cuts determine both the environment indices and the regions
absorbed in an update. Write the environments
as $C_\lambda(a,b)$, $E_\lambda(a,b;\mu)$, and
$aC_\lambda(a,b)$, where $a,b=1,\ldots,\chi$ label the two
renormalized cuts and $\mu=(i,i')$ is a double-layer lattice leg of
dimension $q=D^2$ [Eq.~\eqref{eq:doubletensor}]. The label $\lambda$
specifies an inequivalent cut and its orientation in the tensor cell.
The open $\mu$ leg of $E$ points toward the uncontracted central
network; $C$ and $aC$ have no such leg.

The three absorption regions in Fig.~\ref{fig:hctmrg}(c) contain,
respectively, three $C$, two $E$, and two local double-layer tensors
for a $C$ update; two $E$, one $aC$, and three local tensors for an
$E$ update; and two $C$, one $E$, and one local tensor for an $aC$
update. Internal legs are contracted along the depicted lattice
bonds. Each remaining boundary cut is grouped into a composite
index, with dimension equal to the product of its constituent
dimensions. Denote the enlarged object by $\widetilde X$ for
$X=C,E,aC$. For $E$, each fixed-$\mu$ slice is a matrix on the two
enlarged cuts. With these matrix orientations, compression reads
\begin{equation}
X_{\rm new}^{[\mu]}
=Q_{\rm env}^{(L)}\widetilde X^{[\mu]}P_{\rm env}^{(R)}.
\label{eq:ga_environment_compression}
\end{equation}
The $[\mu]$ label is present only for $E$. The maps act at the two
triangles in Fig.~\ref{fig:hctmrg}(c), leaving this lattice leg
uncompressed. Matching maps are used on adjoining pieces of the
same cut, with their orientations fixed by the geometry.

\subsection{Closed plaquette, matrix reshaping, and synchronous iteration}
\label{app:ga_iteration}

The closed plaquette in Fig.~\ref{fig:hctmrg}(b) contains eight
central double-layer tensors, with eight $E$ and eight $C$ pieces
around them. A convenient index grouping combines one of each into
an octagonal-loop block,
\begin{equation}
\begin{aligned}
&(\mathcal B_\ell)_{(a,\mu),(b,\nu)}\\
&\quad=\sum_{c,\omega}C_\ell(a,c)E_\ell(c,b;\omega)
\mathcal T_{r_\ell}(\mu,\nu,\omega).
\end{aligned}
\label{eq:ga_plaquette_block}
\end{equation}
Here $\ell=1,\ldots,8$ follows the plaquette boundary, and the
local lattice legs are permuted into this order before contraction.
The two matrix indices each pair a $\chi$-dimensional boundary
label with a $q$-dimensional lattice leg. The closed norm is then
$\operatorname{Tr}(\mathcal B_1\cdots\mathcal B_8)$.

For a target cut between blocks $8$ and $1$, opening the opposite
cut between $4$ and $5$ gives
\begin{equation}
L=\mathcal B_5\mathcal B_6\mathcal B_7\mathcal B_8,
\qquad
R=\mathcal B_1\mathcal B_2\mathcal B_3\mathcal B_4.
\label{eq:ga_LR_halves}
\end{equation}
More generally, the open legs are permuted and reshaped so that
$L_{u a}$ has the target composite index $a$ as its column and
$R_{a v}$ has the same index as its row. If this index has dimension
$n$, the matrix shapes are $m_L\times n$ and $n\times m_R$,
where $m_L,m_R$ group the two ends of the opposite cut. The
factorization and maps in Eqs.~\eqref{eq:svdLR} and~\eqref{eq:PQ}
then give $Q_{\rm env}$ of shape $\chi\times n$ and
$P_{\rm env}$ of shape $n\times\chi$.
Only the retained singular values enter the inverse square root;
on their nonzero subspace the paired maps obey
$Q_{\rm env}P_{\rm env}=I_\chi$.
The same ordering of the target constituents is used when applying
the maps in Eq.~\eqref{eq:ga_environment_compression}.

At fixed PEPS tensors, all compression maps and local absorptions
are constructed from the previous environment collection; the
updated $C$, $E$, and $aC$ are then committed simultaneously.

\subsection{Operator insertions, normalization, and cell averages}
\label{app:ga_measurements}

Local observables are evaluated by inserting the operators on their
physical sites in the double-layer network. The same
converged norm environments are used for inserted and uninserted
networks, and each local observable is normalized before taking
its cell average.

Let $c_{a\alpha}$ be the normalized correlator obtained by inserting
$\sigma^\alpha$ on both ends of the $\alpha$ bond from $A_a$ to
its neighbor $B_{b_\alpha(a)}$. With the 16-site labeling of
Appendix~\ref{app:lg}, the exchange energy per site is
\begin{equation}
e_K=-\frac{1}{16}\sum_{a=1}^{8}\sum_{\alpha=x,y,z}
J_\alpha c_{a\alpha}.
\label{eq:ga_exchange_cell_average}
\end{equation}
Listing the bonds from the eight $A$ sites counts each of the
24 cell bonds once, including boundary-crossing bonds, so no
additional factor of $1/2$ is required. Single-site insertions yield
$m_r^\alpha=\langle\sigma_r^\alpha\rangle$. The polarization is their field-weighted average over the 16 sites,
following Eq.~\eqref{eq:k83_field_polarization}, and the total energy
includes the Zeeman contribution $-\sqrt{3}hP$.

For an octagon $p$ of missing color $\gamma_p$, the eight
$\sigma^{\gamma_p}$ operators are inserted simultaneously, with the
minus sign fixed by Eq.~\eqref{eq:Wp}. The mean plaquette flux is
$\overline W=\frac16\sum_{p=1}^6\langle\widehat W_p\rangle$ over the
six representatives in Table~\ref{tab:lg_cell_flux}, each appearing
once per cell. Symmetry-reduced measurements retain the
corresponding multiplicities. The eight-spin expectation is
contracted jointly, not factorized into single-site magnetizations.

\subsection{Initialization dependence in the uniform field}
\label{app:uniform_initialization}

Different PEPS initializations can converge to distinct
variational solutions at the same field. In the saved $D=8$ runs,
the solution continued from zero field has lower energy at
$h/J=0.008$, and the more strongly polarized solution has lower
energy at $0.009$ and $0.010$. Their environment bond dimensions
are $\chi=40$ and $30$, respectively, over these sampled fields.
Both branches were separately checked for convergence with respect
to $\chi$; the observed energy ordering is stable within these checks.
The switch in the selected solution between $0.008$ and $0.009$
produces the plotted steps in polarization and mean flux in
Fig.~\ref{fig:k83_field_tn}(b)--(c).

Writing $\Delta X$ for the more strongly polarized solution minus
the continuation solution at the same field, their energy difference
is $\Delta e=\Delta e_K-\sqrt{3}h\Delta P$.
At $h/J=0.009$, the polarization difference of about $0.3$ lowers
the Zeeman energy of the more strongly polarized solution by
about $5\times10^{-3}J$ per site. The difference in exchange energy
nearly offsets this gain. These comparisons characterize the
finite-$D$ variational solutions obtained from the two initializations.

\section{Third-order effective Majorana models and spectra}
\label{app:field_theory}

\subsection{Oriented paths and the third-order quadratic Hamiltonians}
\label{app:third_order}

Consider an oriented two-bond path
$i\xrightarrow{\alpha}j\xrightarrow{\beta}k$, and let $\gamma$ be
the third bond color at $j$. Define
\begin{equation}
O_{ijk}=\sigma_i^\alpha\sigma_j^\gamma\sigma_k^\beta,
\qquad q_j=\varepsilon_{\alpha\beta\gamma}.
\label{eq:field_path_operator}
\end{equation}
The set $\mathcal P_+$ contains one orientation of each plaquette
corner. We fix it by the central walk
$A_1\to B_1\to A_2\to\cdots\to A_4\to B_4\to A_1$ and
extend this orientation consistently to the other faces.
For the Kekul\'e coloring, the color order reverses between the
two sublattices, giving $q_j=s_j$. For example,
$B_4\to A_1\to B_1$ has $(\alpha,\beta,\gamma)=(y,z,x)$ and
$q_j=+1$, whereas $A_1\to B_1\to A_2$ has $(z,y,x)$ and
$q_j=-1$.

Projecting onto the uniform $\pi$-flux sector and replacing the
virtual-sector energy denominators by common positive excitation
energies gives the path contribution
\begin{equation}
H_{\rm path}^{(3)}\simeq
-\Lambda\sum_{\mathcal P_+}
h_i^\alpha h_j^\gamma h_k^\beta O_{ijk}.
\label{eq:field_path_perturbation}
\end{equation}
Here $h_r^\alpha=h\eta_r$, and $\Lambda$ includes the inverse
energy denominators and time orderings for a path counted once.
The minus sign follows from the three Zeeman factors and two
negative resolvent denominators. This is the local-denominator
approximation to the perturbative projection, rather than an exact
replacement of its resolvents~\cite{Kitaev2006}.
For the staggered field the sublattice factor is
$s_i s_j s_k=s_j$, since the two endpoints belong to the sublattice
opposite to the central site. Writing $\kappa=\Lambda h^3$ in this
orientation convention, the retained path terms are consequently
\begin{equation}
\begin{aligned}
H_{\rm u}^{(3,2)}&=-\kappa\sum_{\mathcal P_+}O_{ijk},\\
H_{\rm s}^{(3,2)}&=-\kappa\sum_{\mathcal P_+}s_jO_{ijk}.
\end{aligned}
\label{eq:field_path_spin_hamiltonians}
\end{equation}
The superscripts specify third order in the field and the part
that is quadratic in matter Majoranas. This quadratic benchmark omits
four-Majorana interactions from other third-order processes and
lower-order exchange renormalizations. The sublattice-sign relation
itself is exact and does not require these approximations.

Use $\sigma_r^\alpha=i b_r^\alpha c_r$,
$u_{rr'}=i b_r^\alpha b_{r'}^\alpha=-u_{r'r}$ on an $\alpha$
bond, and the physical constraint
$b_r^x b_r^y b_r^z c_r=+1$. Pauli--Majorana algebra gives the exact
physical-subspace identity
\begin{equation}
O_{ijk}=i q_j u_{ij}u_{jk}c_i c_k.
\label{eq:field_majorana_path_identity}
\end{equation}
The oriented three-spin reference of Ref.~\cite{HPL2025}, written
with the same path direction, is
\begin{equation}
\begin{aligned}
H_{3{\rm s}}(K)&=-K\sum_{\mathcal P_+}q_j O_{ijk}\\
&=-iK\sum_{\mathcal P_+}u_{ij}u_{jk}c_i c_k.
\end{aligned}
\label{eq:field_oriented_K_reference}
\end{equation}
Fix the same bond gauge $u$ with $W_p=-1$ for both quadratic
models. With $\{c_i,c_k\}=2\delta_{ik}$, define their
single-particle normalization by
\begin{equation}
\begin{aligned}
H_{\rm quad}&=\frac{i}{4}\sum_{r,r'}A_{rr'}c_r c_{r'},
\qquad\mathcal H=iA,\\
\mathcal H(K)&=\mathcal H_0+K\mathsf W.
\end{aligned}
\label{eq:field_single_particle_convention}
\end{equation}
Thus $\mathsf W=\left.\partial_K\mathcal H(K)\right|_{K=0}$ is
dimensionless and is distinct from the plaquette-flux operator.
For a single oriented path its contribution is
$\mathsf W_{ik}=-2i u_{ij}u_{jk}$, with the opposite entry fixed
by Hermiticity. The factor of two accounts for both ordered matrix
entries in $H_{\rm quad}$.

The staggered term in Eq.~\eqref{eq:field_path_spin_hamiltonians}
equals $H_{3{\rm s}}(\kappa)$ because $q_j=s_j$.
The uniform term instead multiplies each reference hopping by
$s_j$. Its endpoints lie on the opposite sublattice:
\begin{equation}
s_i=s_k=-s_j,\qquad \Gamma_{rr'}=s_r\delta_{rr'}.
\label{eq:field_endpoint_sign}
\end{equation}
Replacing the center sign by an endpoint sign therefore yields
\begin{equation}
\begin{aligned}
\delta\mathcal H_{\rm u}^{(3,2)}&=-\kappa\Gamma\mathsf W,\\
\delta\mathcal H_{\rm s}^{(3,2)}&=+\kappa\mathsf W,
\end{aligned}
\label{eq:field_quadratic_mapping}
\end{equation}
as used in Eq.~\eqref{eq:k83_effective_hopping}.
Since these hoppings connect the same sublattice,
$[\Gamma,\mathsf W]=0$, while nearest-neighbor hopping gives
$\{\Gamma,\mathcal H_0\}=0$. Neither identity alone determines
whether the perturbed spectrum has a gap. Reversing the reference
path orientation changes the sign of $\mathsf W$ and the signed
$\kappa$ assigned to the same physical perturbation together.

For completeness, let $\mathcal P$ be a color-preserving spatial
reflection that exchanges $A/B$ and reverses orientation, acting as
$\mathcal P\sigma_r^\alpha\mathcal P^{-1}
=\sigma_{\mathcal P r}^\alpha$.
The uniform Zeeman term preserves $\mathcal P$, whereas the
staggered Zeeman and oriented three-spin terms are odd under it.
The latter preserve $\mathcal P\Theta$, with $\Theta$ the usual
spin time reversal. The spin-sign reflection used for the
three-spin term in Ref.~\cite{HPL2025} is $\mathcal P\Theta$ in
this notation. An unbroken $\mathcal P$ rules out a net chiral
response, but does not forbid a gap; $\mathcal P\Theta$ allows a
chiral response. These symmetry statements do not assign a phase
to the real-field TN states.

\subsection{Window-averaged DOS and twist sampling}
\label{app:dos_sampling}

Figure~\ref{fig:k83_field_dos} uses eigenvalue counts in a symmetric
window $|E|<\epsilon$, without spectral broadening. At fixed
$N$-site supercell, twists assign boundary phases to its
noncontractible cycles. Pairing opposite twists, define
\begin{equation}
d_{b,\epsilon}=
\frac{n_\epsilon(\boldsymbol\theta_b)
+n_\epsilon(-\boldsymbol\theta_b)}{4\epsilon N},
\label{eq:field_dos_pair_count}
\end{equation}
where $n_\epsilon$ counts the eigenvalues of the corresponding
single-particle matrix. Opposite twists are treated as one independent
pair, rather than as two samples. The plotted value is the average of
$d_{b,\epsilon}$ over independent pairs, with standard-error bars.
This agrees with the count divided by $2\epsilon N$ in the figure
caption.

The uniform-type data use $N=2048,4096,8192$, with $B=64$ pairs
at each size. The 2048-site cell follows the supercell construction
of Ref.~\cite{HPL2025}; the larger cells are its index-two and
index-four parity covers. The staggered-type data use
$N=256,512,2048$. Both use $\kappa/J=0.03$ as an independent
effective coupling. The error bars describe twist sampling at a
fixed supercell geometry. Finite-window counts and their low-energy
weight support the spectral comparison in Sec.~\ref{sec:k83_field};
they do not alone determine a CSL or the transition order of the
real Zeeman model.

\end{document}